\documentclass[10pt, draftclsnofoot, onecolumn]{IEEEtran}
\IEEEoverridecommandlockouts

\usepackage{amsmath}
\usepackage{graphicx}
\usepackage{amssymb}
\usepackage{cases}
\usepackage{cite}
\usepackage[ruled,vlined,lined,ruled,linesnumbered]{algorithm2e}
\usepackage{relsize}
\usepackage[nodisplayskipstretch]{setspace}
\usepackage[usenames, dvipsnames]{color}
\usepackage{epstopdf}
\usepackage[font=small,skip=0pt]{caption}
\begin{document}

\title{User Cooperation for Enhanced Throughput Fairness in Wireless Powered Communication Networks}

\author{Mingquan~Zhong,~Suzhi~Bi, and Xiaohui~Lin\\
\thanks{This work was presented in part at the IEEE International Conference on Telecommunications (ICT), Thessaloniki, Greece, May 16-18, 2016.}
College of Information Engineering, Shenzhen University,\\ Shenzhen,  Guangdong, China 518060\\ E-mail:~zhongmingquan@email.szu.edu.cn, \{bsz,~xhlin\}@szu.edu.cn \vspace{-2ex}
       }
\maketitle
\vspace{-5ex}

\begin{abstract}
This paper studies a novel user cooperation method in a wireless powered cooperative communication network (WPCN) in which a pair of distributed terminal users first harvest wireless energy broadcasted by one energy node (EN) and then use the harvested energy to transmit information to a destination node (DN). In particular, the two cooperating users exchange their independent information with each other so as to form a virtual antenna array and transmit jointly to the DN. By allowing the users to share their harvested energy to transmit each other's information, the proposed method can effectively mitigate the inherent user unfairness problem in WPCN, where one user may suffer from very low data rate due to poor energy harvesting performance and high data transmission consumptions. Depending on the availability of channel state information at the transmitters, we consider the two users cooperating using either coherent or non-coherent data transmissions. In both cases, we derive the maximum common throughput achieved by the cooperation schemes through optimizing the time allocation on wireless energy transfer, user message exchange, and joint information transmissions in a fixed-length time slot. We also perform numerical analysis to study the impact of channel conditions on the system performance. By comparing with some existing benchmark schemes, our results demonstrate the effectiveness of the proposed user cooperation in a WPCN under different application scenarios.
\end{abstract}


\IEEEpeerreviewmaketitle

\section{Introduction}
Wireless communication devices are conventionally powered by batteries of limited capacity, which have to be replaced/recharged once the energy is depleted. On one hand, frequent battery replacement/recharging brings a higher operation expense, especially in large-size wireless network, such as wireless sensor networks (WSNs) for environment monitoring. On the other hand, it can cause high probability of communication interrupt that degrades the quality of service. In addition, it could be very inconvenient to replace battery in some special applications (e.g., implanted medical devices). Alternatively, radio frequency (RF) enabled wireless energy transfer (WET) technology allows the wireless devices (WDs) to harvest energy remotely and continuously from the RF signals radiated by some dedicated energy nodes (ENs) \cite{2014:Krikidis,2014:Bi,2015:Lu,2016:Bi,2013:Zhou,2016:JXU}. Compared to the conventional battery-powered communications, WET can effectively reduce the network maintenance cost and also provide more stable services.

One interesting application of WET is wireless powered communication network (WPCN), where WDs transmit information using the energy harvested by means of WET \cite{2014:Ju1,2014:Liu2,2014:Ju2,2014:Huang1,2014:Lee1,{2015:Bi:Placement Optimization},2016:BIZHANG,2015:Che,2015:Nasir,2014:Krikidis,2014:Ju3,2015:HeCHen}. For instance, \cite{2014:Ju1} proposed a harvest-then-transmit protocol in WPCN, where one hybrid access point (HAP) with single-antenna first broadcasts RF energy to all users in the downlink, and then the users perform wireless information transmission (WIT) to the HAP in the uplink using their individually harvested energy in a time-division-multiple-access (TDMA) manner. \cite{2014:Liu2} extended the single-antenna HAP in \cite{2014:Ju1} to a multi-antenna HAP that enables more efficient energy transmission via energy beamforming, and more spectrally efficient SDMA (space division multiple access) based information transmission as compared to TDMA. Moreover, \cite{2014:Ju2} considered using full-duplex HAP which can transmit energy and receive user data simultaneously with advanced self-interference cancelation techniques. Despite of their different settings, all the above schemes consider using a HAP for both transmitting energy and receiving information. Although enjoying lower deployment and production cost than using a pair of separated energy and information access points (APs), the use of HAP can induce serious user unfairness, named doubly-near-far problem, where users far away from the HAP achieve very low throughput because they suffer from both poor energy harvesting performance and high data transmission consumptions \cite{2014:Ju1}.

To enhance user fairness, \cite{2014:Ju3} proposed a two-user cooperation scheme where the near user helps relay the far user's information to the HAP. \cite{2015:HeCHen} extended \cite{2014:Ju3} to a multi-relay scenario and proposed a harvest-then-cooperate protocol to coordinate the transmissions of nearby users to forward the message of a far-away user. Both works consider using a HAP to transmit energy and receive information, which, however, is the essential cause of the doubly-near-far problem. To further enhance user fairness, separately located energy and information APs are considered to more flexibly balance the energy and information transmissions, as now the poor energy harvesting performance of a WD can be compensated by low information transmit power to a nearby information AP \cite{2014:Huang1,2014:Lee1,{2015:Bi:Placement Optimization},2016:BIZHANG}.

In this paper, we present a new user cooperation method in WPCN with separately located energy node (EN) and information AP to enhance user fairness performance. As shown in Fig.~1, the two energy-harvesting users $X$ and $Y$ exchange their individual messages with each other to form a virtual antenna array, and transmit jointly to the destination node (DN). The cooperative transmission of the two users can be performed in either coherent or non-coherent manner. In particular, coherent transmission using distributed transmit beamforming (DTB) attains the maximum signal-to-noise power ratio  (SNR), which however requires accurate knowledge of wireless channel state information at transmitter side (CSIT) for achieving high distributed beamforming gain. In practice, the acquisition of highly accurate CSIT in WPCN may degrade the overall communication performance due to the time and energy consumed on the receiver side CSI feedback. On the other hand, sub-optimal non-coherent transmission can be performed, e.g., using space-time block codes (STBC) \cite{2003:Laneman}, without the knowledge of CSIT. In this paper, we consider both cases under different CSIT availability conditions.

The key contributions of this paper are summarized as follows:
\begin{itemize}
  \item We present a new user cooperation method for enhancing the throughput fairness in WPCN. Specifically, a pair of wireless powered terminal users first exchange their independent messages with each other and then transmit jointly to the DN. Compared to the existing cooperation scheme where one user acts as the relay for the other, the proposed user cooperation method allows the two users to share their harvested energy and to transmit jointly, thus achieving both energy diversity and channel diversity gains.
  \item With both coherent and non-coherent cooperative transmissions employed by the two users, we derive the maximum common throughput achieved by the cooperation scheme through optimizing the time allocation on wireless energy transfer, user message exchange, and joint information transmissions in a fixed time slot under different transmission schemes and decoding ability.
  \item We also perform numerical analysis to study impact of  system setups to the performance of the proposed user cooperation method. Through comparisons with other benchmark schemes, we show that the proposed cooperation can effectively improve the throughput performance, especially when the inter-user channels are sufficiently strong to support efficient information exchange and the two users have comparable user-to-DN channels.

\end{itemize}

The rest of this paper is organized as follows: Section II presents the system model. We formulate the optimal user cooperation problem in Section III. Section IV and Section V study the optimal time allocation to maximize the common throughput of the proposed cooperation method using non-coherent and coherent transmissions, respectively. Some benchmark methods are introduced in Section VI. We present simulation results in Section VII. Finally, Section VIII concludes the paper.

\begin{figure}
  \centering
   \begin{center}
      \includegraphics[width=0.7\textwidth]{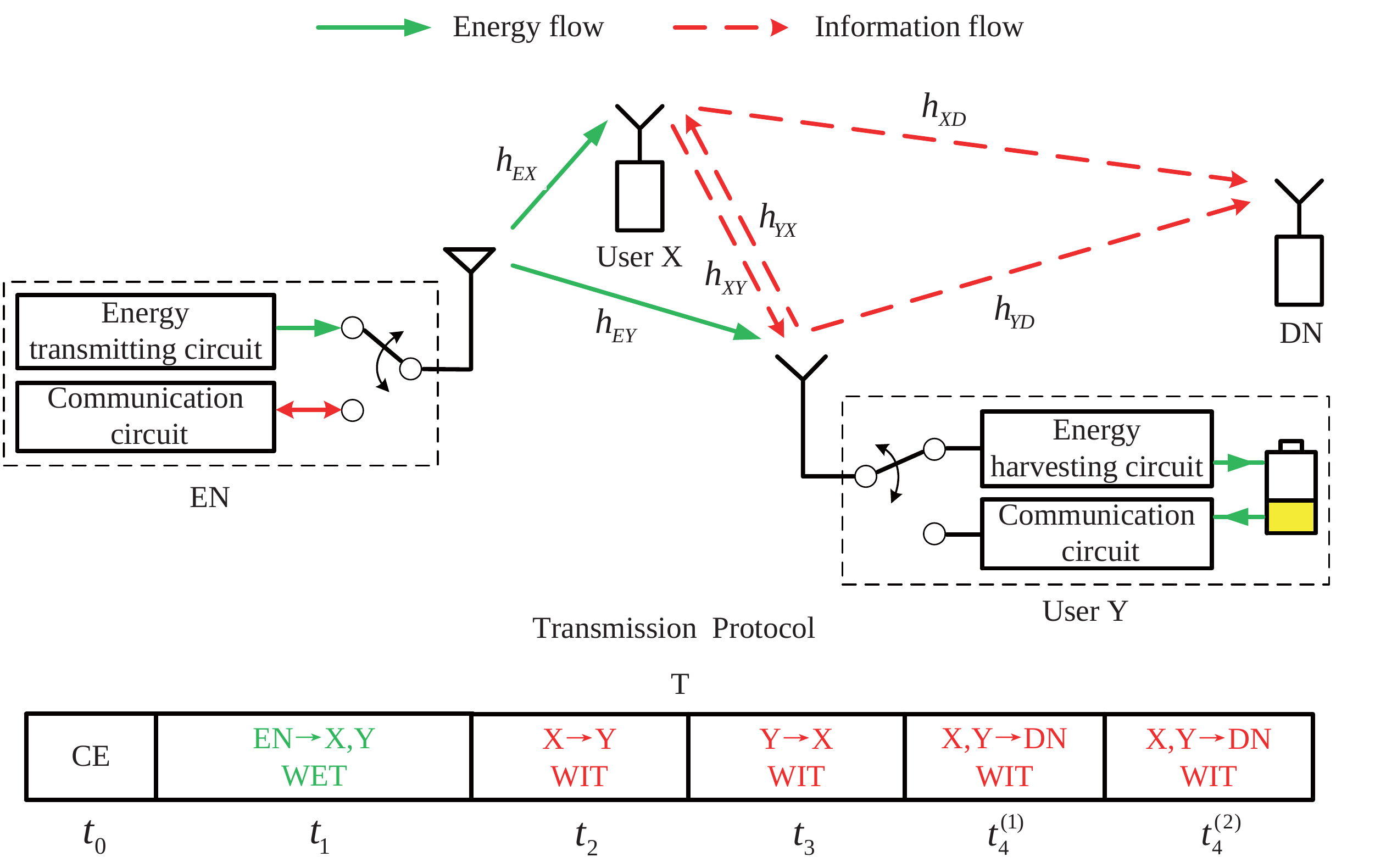}
   \end{center}
  \caption{The proposed user cooperation method and the operating protocol. }
  \label{Fig.1}
\end{figure}
\section{System Model}
\subsection{Transmission Protocol}

As shown in Fig.~1, we consider a WPCN with two users $X$ and $Y$ who first harvest RF energy from the EN and then transmit cooperatively their data to the DN. The EN is assumed to have a constant energy supply and both terminal users have no other embedded energy source, thus need to store the harvested energy in a rechargeable battery for information transmission to the DN. It is assumed that each node is equipped with single antenna. For each user, the antenna is used for both energy harvesting and communication in a time-division-duplexing (TDD) manner \cite{2013:Zhou} (e.g, the circuit structure of user $Y$ is illustrated in Fig.~1). The EN also has a similar TDD circuit structure to switch between energy transfer and communication with the WDs. Notice that the communication circuit of the EN is only for performing channel estimation (CE), rather than transmitting/receiving user data to/from the WDs.

It is assumed that all the channels are under quasi-static flat-fading, where the channel gains remain constant during each transmission block of duration $T$ but vary from one block to another. At the beginning of a transmission block, CE is performed within a fixed duration $t_0$. Then, in the remainder of a tagged transmission block, $t_1$ amount of time is assigned for WET while the remaining time is assigned for WIT. In the next two time slots with duration $t_2$ and $t_3$, respectively, the two users exchange with each other their own messages. In the last time slot of length $t_4$, the two users transmit jointly their information to the DN. Specially, $t_4^{(1)}$ amount of time is allocated to transmit user $X$'s information, and the rest of the time slot with duration $t_4^{(2)}$ is for transmitting $Y$'s information, with $t_4=t_4^{(1)}+t_4^{(2)}$. Note that we have a total time constraint
\begin{equation}\label{T}
t_0+t_1+t_2+t_3+t_4^{(1)}+t_4^{(2)}=T.
\end{equation}
For the simplicity of exposition, we set without loss of generality $T=1$ throughout this paper.

\subsection{Channel Estimation Methods}
The notations of channel gains are shown in Fig.~1. In the CE stage, user $X$ and $Y$ take turns to broadcast their pilot signals, so that EN has the knowledge of $h_{EX}$ and $h_{EY}$, the DN knows $\alpha_{_{XD}}$ and $\alpha_{_{YD}}$, and user $X$ ($Y$) knows $\alpha_{_{YX}}$ ($\alpha_{_{XY}}$), respectively, where $\alpha_{_{XY}}$ denotes the complex channel coefficient between $X$ and $Y$ with $h_{XY}\triangleq |\alpha_{XY}|^2$. Then, each node feeds back their known CSI to a control point, which calculates and broadcasts the optimal time allocation $(t_1^*,t_2^*,t_3^*,t_4^{(1)*},t_4^{(2)*})$ to all the nodes in the network. At the end of each time slot, we assume that user $X$ and $Y$ reserve a fixed amount of energy for performing CE in the next time slot. Notice that under this basic setup, user $X$ and $Y$ have no knowledge of user-to-DN channel, i.e., $\alpha_{_{XD}}$ and $\alpha_{_{YD}}$, thus coherent transmit beamforming is not applicable at user $X$ and $Y$. In this case, non-coherent STBC is applied by the two users to transmit cooperatively to the DN. However, if the central point or DN feeds back additional $\alpha_{_{XD}}$ and $\alpha_{_{YD}}$ to user $X$ and $Y$, respectively, coherent transmission can be applied by the two users to enhance the communication performance. Depending on the availability of CSIT at the two users, we introduce the following two cooperation methods.
\subsection{STBC-based Cooperation}
We first consider the case that the two users perform non-coherent STBC-based cooperation. In the first time slot, we let $P_t$ denote the fixed transmission power of the EN and assume that the energy harvested from noise is negligible by the users. Then, the amount of energy harvested by user $X$ and $Y$ can be expressed as \cite{2014:Ju1}
\begin{equation}\label{energy}
  E_X={\eta}{t_1}{P_t}{h_{EX}},\  E_Y={\eta}{t_1}{P_t}{h_{EY}},
\end{equation}
where $0<\eta<1$ denotes the energy harvesting efficiency assumed equal for both users.

After harvesting wireless energy from the EN, the two users exchange their independent information with each other and then transmit jointly to DN. Here, we assume that both user $X$ and user $Y$ exhaust the harvested energy for transmitting information, and their transmit power is constant during the WIT stage. Then, the transmit power of $X$ and $Y$ is $P_X=E_X/(t_2+t_4)$ and $P_Y=E_Y/(t_3+t_4)$, respectively. Let $S_X(t)$ denote the transmitted baseband signal of user $X$ in $t_2$ with $ E[|S_X(t)|^2] = 1$, the received signal at user $Y$ is then expressed as
\begin{equation}
Z_Y^{(2)}(t)=\sqrt{P_X}{\alpha}_{_{XY}}S_X^{(2)}(t)+n_Y^{(2)}(t),
\end{equation}
where $t\in(t_0+t_1,t_0+t_1+t_2]$, and $n_Y^{(2)}(t)$ denotes the receiver noise at user $Y$. Without loss of generality, we assume that the receiver noise power is $N_0$ at all receivers. Then, user $Y$ can decode the $X$'s information at a rate given by \footnote{We do not consider the energy consumption for information decoding in this paper.}
\begin{equation}
\label{Rx2}
R_X^{(2)}=t_2\log_{2}\left(1+\frac{E_Xh_{XY}}{(t_2+t_4)N_0}\right).
\end{equation}

At the same time, the DN also receives the information broadcasted by user $X$ during $t_2$ due to the broadcasting feature of wireless communication. The channel power gains from user $X$ to DN and the receiver noise at the DN are denoted by $h_{XD}$ and $n_D^{(2)}(t)$, respectively. Then, the DN receives
\begin{equation}
\label{Zd2}
Z_D^{(2)}(t)=\sqrt{P_X}{\alpha}_{_{XD}}S_X^{(2)}(t)+n_D^{(2)}(t),
\end{equation}
where $t\in(t_0+t_1,t_0+t_1+t_2]$.

Similarly, let $S_Y(t)$ denote the transmitted baseband signal of the user $Y$ in $t_3$ with $ E[|S_Y(t)|^2] =1$. The signals received by user $X$ and the DN during $t_3$, are expressed as
\begin{equation}
Z_X^{(3)}(t)=\sqrt{P_Y}{\alpha}_{_{YX}}S_Y^{(3)}(t)+n_X^{(3)}(t),
\end{equation}
\begin{equation}
\label{Zd3}
Z_D^{(3)}(t)=\sqrt{P_Y}{\alpha}_{_{YD}}S_Y^{(3)}(t)+n_D^{(3)}(t),
\end{equation}
where $t\in(t_0+t_1+t_2,t_0+t_1+t_2+t_3]$ and the data rate from user $Y$ to $X$  in duration $t_3$ is given by
\begin{equation}
\label{Ry3}
R_Y^{(3)}=t_3\log_{2}\left(1+\frac{E_Yh_{YX}}{(t_3+t_4)N_0}\right).
\end{equation}

In the $4$-th time slot, the two users use Alamouti STBC transmit diversity scheme \cite{Alamouti} for joint information transmission with $t_4^{(1)} = t_4^{(2)}$. Accordingly, the received SNR at the DN for both users is
\begin{equation}
\gamma_{Z_D^{(4)}(t)}= \frac{P_Xh_{XD}+P_Yh_{YD}}{N_0},
\end{equation}  where $h_{XD} \triangleq {|{\alpha}_{_{XD}}|^2}$ and $h_{YD} \triangleq {|{\alpha}_{_{YD}}|^2} $.

Notice that the DN can overhear the transmission of user $X$ ($Y$) in the $2$-nd ($3$-rd) time slot, although not dedicated to it. In theory, the DN can improve the data rates of both users with the overheard signals, which, however, requires a well-designed coding scheme and adequate hardware complexity at the DN to perform joint decoding \cite{2014:Ju3}. When a simple coding scheme (or simple DN receiver structure) is used, such that the DN only decodes each user's information transmission in the $4$-th time slot, the achievable data rate from user $X$ to DN is
\begin{equation}
\label{Rx4}
R_X^{(4)}=t_4/2\log_{2}\left(1+ \frac{E_Xh_{XD}}{(t_2+t_4)N_0} + \frac{E_Yh_{YD}}{(t_3+t_4)N_0} \right ),
\end{equation} and $R_Y^{(4)}=R_X^{(4)}$ for user $Y$. Otherwise, if DN can jointly decode the information transmitted by the users, say user $X$ in $t_2$ and $t_4$, the achievable rates from $X$ and Y to DN are
\begin{equation}\label{Rx4j}
R_X^{(4)}=t_4/2\log_{2}\left(1+ \frac{E_Xh_{XD}}{(t_2+t_4)N_0} + \frac{E_Yh_{YD}}{(t_3+t_4)N_0} \right )
+ t_2\log_{2}\left(1+\frac{E_Xh_{XD}}{(t_2+t_4)N_0}\right),
\end{equation}

\begin{equation}\label{Ry4j}
R_Y^{(4)}= t_4/2\log_{2}\left(1+ \frac{E_Xh_{XD}}{(t_2+t_4)N_0} + \frac{E_Yh_{YD}}{(t_3+t_4)N_0} \right )
+ t_3\log_{2}\left(1+\frac{E_Yh_{YD}}{(t_3+t_4)N_0}\right).
\end{equation}

\subsection{DTB-based Cooperation}
When $\alpha_{XD}$ and $\alpha_{YD}$ are known at $X$ and $Y$, respectively, DTB-based cooperation can be performed by the two users. Similarly, the harvested energy in $t_1$ and the information transmission in $t_2$ and $t_3$ follow (\ref{energy}), (\ref{Rx2}), and (\ref{Ry3}), respectively. In the $4$-th time slot, however, $X$ and $Y$ use $\alpha_{XD}^*/|\alpha_{XD}|$ and $\alpha_{YD}^*/|\alpha_{YD}|$ as the beamforming vectors, respectively, and the received SNR at the DN for the two users are
\begin{equation}
\gamma_{Z_D^{(41)}(t)}= \gamma_{Z_D^{(42)}(t)} = \frac {({\sqrt{P_Xh_{XD}} + \sqrt{P_Yh_{YD}}})^2} {N_0}.
\end{equation}

Notice that, unlike the STBC-based cooperation with equal time allocation between the two users, the time durations to transmit the two users' data can be different for the DTB-based scheme, i.e., $t_4^{(1)}\neq t_4^{(2)}$ in general. When the DN only decodes the users' information in the 4-th time slot, the achievable rates of the two users are
\begin{equation}\label{Rx4m}
  R_X^{(4)}=t_4^{(1)}\log_{2} \left(1+{\left({\sqrt{\frac{E_Xh_{XD}}{(t_2+t_4)N_0}} + \sqrt{\frac{E_Yh_{YD}}{(t_3+t_4)N_0}}}\right)^2} \right),
\end{equation}
\begin{equation}\label{Ry4m}
  R_Y^{(4)}=t_4^{(2)}\log_{2} \left(1+{\left({\sqrt{\frac{E_Xh_{XD}}{(t_2+t_4)N_0}} + \sqrt{\frac{E_Yh_{YD}}{(t_3+t_4)N_0}}}\right)^2} \right).
\end{equation}

Otherwise, the achievable rates of user X and Y when joint decoding is applied at the DN are
\begin{equation}\label{Rx4mj}
  \begin{split}
    R_X^{(4)} & = t_4^{(1)}\log_{2} \left(1+{\left({\sqrt{\frac{E_Xh_{XD}}{(t_2+t_4)N_0}} + \sqrt{\frac{E_Yh_{YD}}{(t_3+t_4)N_0}}}\right)^2} \right)
      + t_2\log_{2}\left(1+\frac{E_Xh_{XD}}{(t_2+t_4)N_0}\right),
  \end{split}
\end{equation}
\begin{equation}\label{Ry4mj}
  \begin{split}
    R_Y^{(4)} & = t_4^{(2)}\log_{2} \left(1+{\left({\sqrt{\frac{E_Xh_{XD}}{(t_2+t_4)N_0}} + \sqrt{\frac{E_Yh_{YD}}{(t_3+t_4)N_0}}}\right)^2} \right)
      + t_3\log_{2}\left(1+\frac{E_Yh_{YD}}{(t_3+t_4)N_0}\right).
  \end{split}
\end{equation}

\section{Problem Formulation}
From the discussions in Section II, the achievable rates of user $X$ and $Y$ are
\begin{equation}\label{1}
  R_X = \min \left(R_X^{(2)},R_X^{(4)}\right),\ R_Y = \min \left(R_Y^{(3)},R_Y^{(4)}\right),
\end{equation}
where $R_X^{(2)}$ and $R_Y^{(3)}$ are in (\ref{Rx2}) and (\ref{Ry3}), respectively. For $R_X^{(4)}$ and $R_Y^{(4)}$, their expressions depend on the specific transmitter and receiver structures used in the following schemes:

\begin{itemize}
  \item (STBC-NJD): STBC-based user cooperation and DN only decodes user message in the $4$-th time slot, i.e., no joint decoding capability of user message across different time slots. In this case, $R_Y^{(4)}=R_X^{(4)}$ in (\ref{Rx4}).
  \item (STBC-JD): STBC-based user cooperation and DN with joint decoding capability, i.e., $R_X^{(4)}$ in (\ref{Rx4j}) and $R_Y^{(4)}$ in (\ref{Ry4j}).
  \item (DTB-NJD): DTB-based cooperation and DN without joint decoding capability, i.e., $R_X^{(4)}$ in (\ref{Rx4m}) and $R_Y^{(4)}$ in (\ref{Ry4m}).
  \item (DTB-JD): DTB-based cooperation and DN with joint decoding capability: $R_X^{(4)}$ in (\ref{Rx4mj}) and $R_Y^{(4)}$ in (\ref{Ry4mj}).
\end{itemize}

In WPCNs, the user data rates can differ significantly, e.g., by two-orders of amplitude, because of the disparities in both energy harvesting performance and information transmit power consumptions. As a common indicator of throughput fairness, we adopt the minimum throughput of the two users as the performance metric \cite{2014:Ju1}, i.e.,
\begin{equation}
\label{Rmaxmin}
R=\min(R_X,R_Y).
\end{equation}
In particular, we are interested in the following optimal time allocation problem to maximize the throughput
\begin{equation}
   \begin{aligned}
    &\max_{t_1,t_2,t_3,t_4} & &  \min(R_X,R_Y)\\
    &\text{s. t.}    & & t_1+t_2+t_3+t_4 = 1 - t_0, \label{P1}\\
    & & & t_1,t_2,t_3,t_4\geq 0,
   \end{aligned}
\end{equation}
where $R_X$ and $R_Y$ are in (\ref{1}).

By introducing an auxiliary variable $z$, problem (\ref{P1}) can be equivalently written as
\begin{equation}
\label{2}
   \begin{aligned}
    &\max_{z,t_1,t_2,t_3,t_4} & &  z\\
    &\text{s. t.}    & & R_X^{(2)} \geq z, R_X^{(4)} \geq z, R_Y^{(3)} \geq z, R_Y^{(4)} \geq z, \\
    & & & t_1+t_2+t_3+t_4 = 1 - t_0,\\
    & & & t_1,t_2,t_3,t_4\geq 0,
   \end{aligned}
\end{equation}
which is a non-convex problem, because for any of the four transmission schemes in consideration, neither one of $\left\{R_X^{(2)},R_X^{(4)},R_Y^{(3)},R_Y^{(4)}\right\}$ is jointly concave in $\left(t_1,t_2,t_3,t_4\right)'$. Therefore, the optimal solution of (\ref{2}) cannot be efficiently obtained using standard convex optimization technique, such as interior point method. Besides, effective transformation of (\ref{P1}) to an equivalent convex form is currently absent. However, we show in the following sections that the optimal solution to (\ref{P1}) can be obtained by exploiting the monotonic properties of the time allocation solutions.

\section{Throughput Performance of Non-coherent User Cooperation}
In this section, we study the optimal throughput performance of the considered user cooperation scheme when the two users apply STBC to perform joint transmission to the DN, i.e., the STBC-NJD and STBC-JD schemes. For the STBC-NJD scheme, we first analyze in Section IV.A the properties of an optimal solution to (\ref{P1}). Based on the analysis, we then propose an efficient algorithm to solve (\ref{P1}) optimally when STBC-NJD scheme is used. In addition, we also derive an achievable throughput when STBC-JD scheme is used.

\subsection{Analysis of STBC-NJD Scheme}
We first consider the case where the two users transmit jointly to the DN using STBC and the DN can only decode each user' message in the $4$-th time slot. To begin with, we first show that the optimal solution to (\ref{P1}) should allow the two terminal users to transmit at an equal rate, i.e., $R^*_X =R^*_Y$. Otherwise, if $R_X^* \neq R_Y^*$, we assume without loss of generality that $R_X >R_Y$, and the case with $R_X<R_Y$ follows. In this case, $R_X^{(2)}>R_Y^{(3)}$ must hold because $R_X^{(4)} = R_Y^{(4)}$ due to the STBC-based cooperation in the $4$-th time slot. Note that given a pair of $(t_1,t_4)$, $R_X^{(2)}$ ($R_Y^{(3)}$) is an increasing (a decreasing) function of $t_2$, for $t_2 \in\left[1-t_0-t_1-t_4\right]$, which is proved in Lemma 4.1 and demonstrated numerically in Fig.~2. Accordingly, we can always adjust $t_2$, and thus $t_3$, to improve the objective of (\ref{P1}) until $R_X^{(2)}=R_Y^{(3)}$. Therefore, we can conclude that $R_X^* =R_Y^*$ must hold for the optimal solution of (\ref{P1}). Accordingly, the optimal throughput in (\ref{P1}) is often referred to as \emph{common throughput} \cite{2014:Ju1}.
\begin{figure}
  \centering
   \begin{center}
      \includegraphics[width=0.65\textwidth]{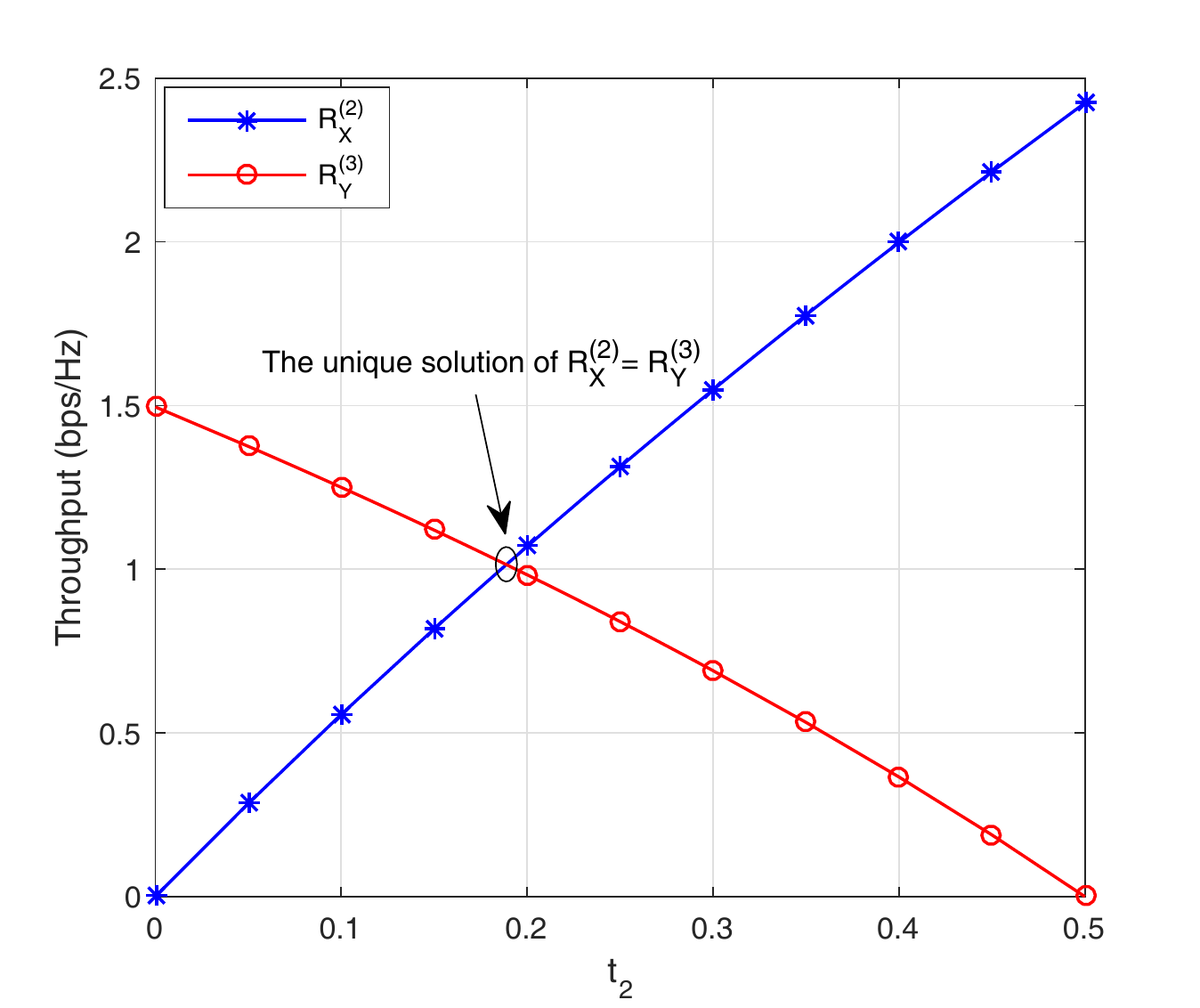}
   \end{center}
  \caption{Numerical results of the monotonic properties in Lemma 4.1. Without loss of generality, the system parameters are given in the simulation section. }
  \label{Fig.2}
\end{figure}

\underline{\emph{Lemma}} \emph{4.1}: $R_X^{(2)}$ increases monotonically and $R_Y^{(3)}$ decreases monotonically in $t_2 \in\left[0,T_0\right]$, where $T_0= t_2 +t_3$ is a fixed parameter.

\emph{Proof:}
Please refer to Appendix A.

Then, we show that $R_X^{(2)*}=R_X^{(4)*}$. Otherwise, if $R_X^{(2)*}>R_X^{(4)*}$ (or $R_X^{(2)*}<R_X^{(4)*}$), we can easily increase the objective in (\ref{P1}) by allocating more time on WET, and less time for user cooperation in $t_2$ and $t_3$ (or joint transmission in $t_4$). Similarly, we have $R_Y^{(3)*}=R_Y^{(4)*}$. Based on the above analysis, we conclude that the optimal solution must satisfy
\begin{equation}\label{P2}
  R_X^{(2)*}=R_Y^{(3)*}=R_X^{(4)*}.
\end{equation}
Besides, $R_X^{(4)*}=R_Y^{(4)*}$ holds because of the Alamouti STBC in use. We can express the terms in (\ref{P2}) as functions of time allocation as following
\begin{align}
R_X^{(2)}&=t_2\log_{2}\left(1+ \rho_1 \frac{t_1}{t_2+t_4}\right) ,  \label{rx2} \\
R_Y^{(3)}&=t_3\log_{2}\left(1+ \rho_2 \frac{t_1}{t_3+t_4}\right) ,   \label{ry3}\\
R_X^{(4)}&=t_4/2\log_{2}\left(1+ \rho_3 \frac{t_1}{t_2+t_4}+ \rho_4 \frac{t_1}{t_3+t_4}\right) ,\label{rx4}
\end{align}
where ${\rho_1}\triangleq{ \eta P_t h_{EX} h_{XY} } / {N_0}$, ${\rho_2}\triangleq{ \eta P_t h_{EY} h_{YX} } / {N_0}$, ${\rho_3}\triangleq{ \eta P_t h_{EX} h_{XD}}/ {N_0}$, and ${\rho_4}\triangleq{ \eta P_t h_{EY} h_{YD}}/ {N_0}$.

\begin{figure}
  \centering
   \begin{center}
      \includegraphics[width=0.65\textwidth]{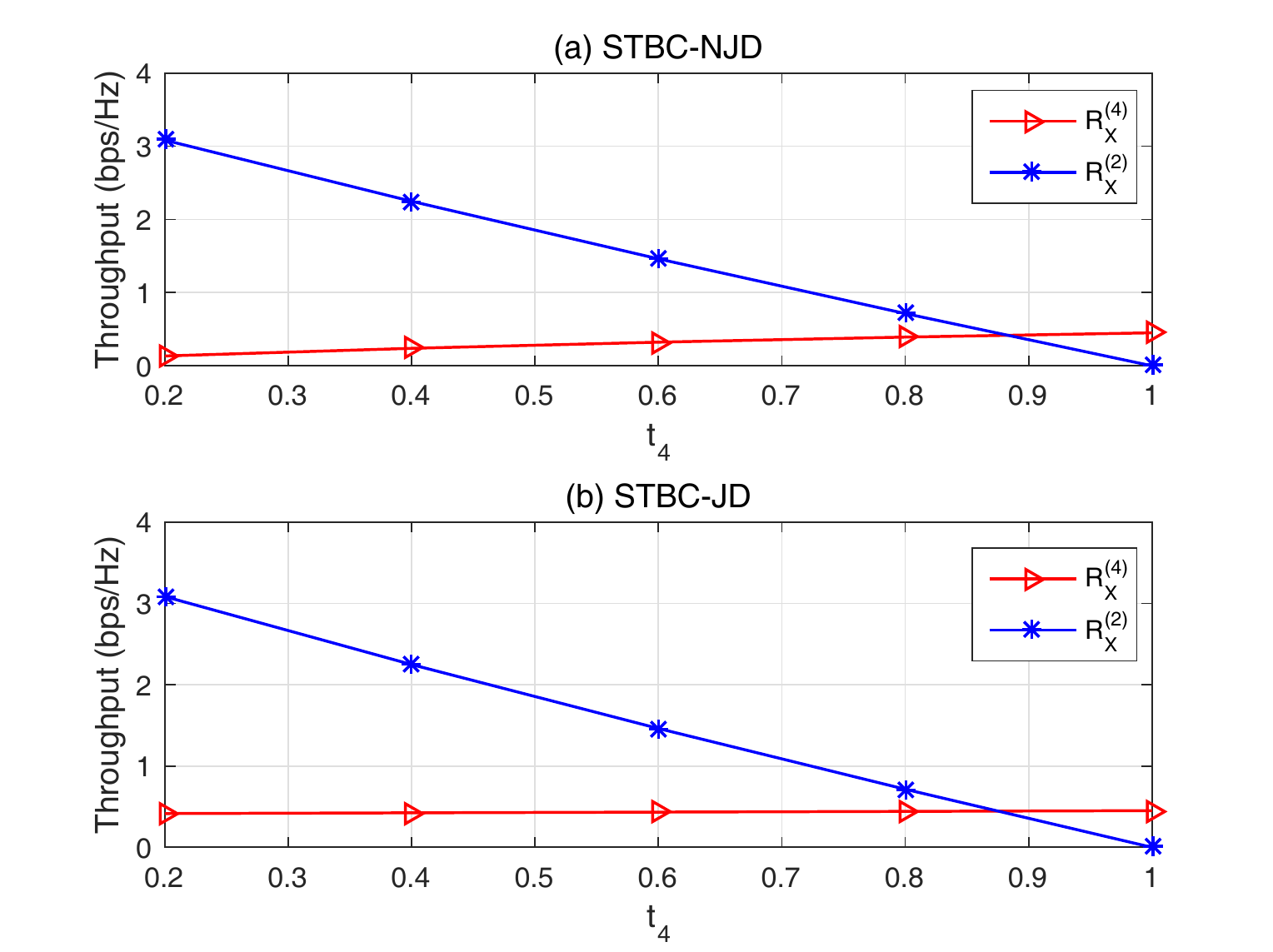}
   \end{center}
  \caption{Numerical results of the monotonic properties in Lemma 4.2 based on Alamouti code}
  \label{Fig.3}
\end{figure}

\subsection{Optimal Solution Algorithm to the STBC-NJD Scheme}
Based on the above analysis, we then propose an efficient algorithm to solve (\ref{P1}). To begin with, we show that there always exists a unique time allocation $(t_2,t_3,t_4)$ that satisfies (\ref{P2}) given a fixed $t_1$. To see this, from Lemma 4.1, we can always find a unique set of $(t_2,t_3)$ to satisfy $R_X^{(2)} = R_{Y}^{(3)}$ when a set of $(t_1,t_4)$ is given, such that $t_2+t_3 = 1-t_0-t_1-t_4$ is a fixed parameter. Equivalently, we can denote $R_X^{(2)}$ and $R_Y^{(3)}$ as functions of $(t_1,t_4)$, i.e., $R_X^{(2)}(t_1,t_4)$ and $R_Y^{(3)}(t_1,t_4)$, respectively. Besides, $R_X^{(4)}$ can also be expressed as a function $(t_1,t_4)$, i.e., $R_X^{(4)}(t_1,t_4)$, because $(t_2,t_3)$ is uniquely determined by a pair of $(t_1,t_4)$. Then, given a fixed $t_1$, a unique $t_4$ can be found to satisfy $R_X^{(2)}(t_1,t_4) = R_X^{(4)}(t_1,t_4)$, because $R_X^{(2)} = 0$ when $t_4 = 1-t_0-t_1$ and $R_X^{(2)}$ decreases with $t_4 \in\left[0,1-t_0-t_1\right]$, while $R_X^{(4)} = 0$ when $t_4 = 0$ and $R_X^{(4)}$ increases with $t_4 \in\left[0,1-t_0-t_1\right]$. In particular, given a fixed $t_1$, the monotonic properties of $R_X^{(2)}(t_1,t_4)$ and $R_X^{(4)}(t_1,t_4)$ as a function of $t_4$ are proved in the following Lemma 4.2 and illustrated numerically in Fig.~\ref{Fig.3}.

\underline{\emph{Lemma}} \emph{4.2}: $R_X^{(4)}$ increases monotonically and $R_X^{(2)}$ decreases monotonically in $t_4\in \left[0,T_1\right]$, where $T_1=t_2+t_3+t_4$ is a fixed parameter.

\emph{Proof:}
Please refer to Appendix B.

\begin{algorithm}
\footnotesize
 \SetAlgoLined
 \SetKwData{Left}{left}\SetKwData{This}{this}\SetKwData{Up}{up}
 \SetKwRepeat{doWhile}{do}{while}
 \SetKwFunction{Union}{Union}\SetKwFunction{FindCompress}{FindCompress}
 \SetKwInOut{Input}{input}\SetKwInOut{Output}{output}
 \Input{time duration $T=1$, channel estimation time $t_0$}
 \Output{the optimal time allocation of $\left\{t_1^*,t_2^*,t_3^*,t_4^*\right\}$}
 Initialize: $t_1\leftarrow 0$, $R^*\leftarrow 0$, $\Delta \leftarrow$ small positive step size\;
   \While{$t_1 \leq 1-t_0$}{
 $t_1 \leftarrow t_1+\Delta$\;
    $UB_4 \leftarrow  1-t_0-t_1$, $LB_4 \leftarrow 0$\;
    \Repeat{$|R_X^{(2)}-R_X^{(4)}|<\sigma$}{
    $t_4\leftarrow \left(UB_4+LB_4\right)/2$\;
    $UB_2 \leftarrow  1 - t_0 - t_1 -t_4$, $LB_2 \leftarrow 0$\;
    \Repeat{$|R_X^{(2)}-R_Y^{(3)}|<\sigma$}{
    $t_2\leftarrow \left(UB_2+LB_2\right)/2$\;
    $t_3\leftarrow 1-t_0-t_1-t_4-t_2$\;
    Calculate $R_X^{(2)}$ and $R_Y^{(3)}$ using (\ref{rx2}) and (\ref{ry3}), respectively\;
    \eIf{$R_X^{(2)}>R_Y^{(3)}$}{
    $UB_2 \leftarrow t_2$;}
    {
    $LB_2 \leftarrow t_2$;
    }
    }

    Given $t_2,t_3,t_4$, calculate $R_X^{(4)}$ using (\ref{rx4})\;
    \eIf{$R_X^{(4)}>R_X^{(2)}$}{
    $UB_4 \leftarrow t_4$;}
    {
    $LB_4 \leftarrow t_4$;
    }
    }
    $R\leftarrow \min\left(R_X^{(2)},R_X^{(4)}\right)$\;
    \If{$R>R^*$}{
    $R^*\leftarrow R$, \ \
    $\left\{t_1^*,t_2^*,t_3^*,t_4^*\right\}\leftarrow \left\{t_1,t_2,t_3,t_4 \right\}$\;
    }

}
\textbf{Return} $\left\{t_1^*,t_2^*,t_3^*,t_4^*\right\}$.
\caption{Optimal solution to (\ref{P1}) for the STBC-NJD scheme}
\label{alg1}
\end{algorithm}

Now that a unique time allocation $(t_2,t_3,t_4)$ can be found with a fixed $t_1$, the optimal solution to (\ref{P1}) can be obtained by a simple line search over $t_1\in \left[0,1-t_0\right]$. A pseudo-code of the above searching algorithm is summarized in Algorithm 1, where the lines $7-17$ correspond to the bi-section search over $t_2$ and lines $4-24$ correspond to the bi-section search over $t_4$. The time complexity of the algorithm is proportional to $1/\Delta \cdot \left[\log(1/\sigma)\right]^2$, where $\Delta$ and $\sigma$ are small positive parameters determined by the solution precision requirement. The proposed algorithm is of low complexity, which enables fast calculation of the optimal time allocation solution.

\subsection{Achievable Throughput of STBC-JD Scheme}
When the DN can jointly decode each user's message transmitted across two different time slots and based on different encoding methods, the achievable rates of the two users in the $4$-th time slot are:
\begin{equation}\label{rx4j}
 \begin{split}
     R_X^{(4)}&= t_4/2\log_{2}\left(1+ \rho_3 \frac{t_1}{t_2+t_4}+ \rho_4 \frac{t_1}{t_3+t_4}\right)
       +t_2\log_{2}\left(1+ \rho_3 \frac{t_1}{t_2+t_4}\right),
   \end{split}
\end{equation}
\begin{equation}\label{ry4j}
 \begin{split}
     R_Y^{(4)}&= t_4/2\log_{2}\left(1+ \rho_3 \frac{t_1}{t_2+t_4}+ \rho_4 \frac{t_1}{t_3+t_4}\right)
       +t_3\log_{2}\left(1+ \rho_4 \frac{t_1}{t_3+t_4}\right).
   \end{split}
\end{equation}
In this case, however, $R_X$ and $R_Y$ are different in general. This is because, to achieve $R_X^{(2)}=R_X^{(3)}$, we have shown that a pair of $(t_2,t_3)$ is uniquely determined given a fixed $t_4$. However, by substituting such a pair of $(t_2,t_3)$ into (\ref{rx4j}) and (\ref{ry4j}), we can see that $R_X^{(4)}\neq R_Y^{(4)}$ in general, because of the inherent difference of the user-to-DN and EN-to-user channels between the two users. Due to the non-convex nature of problem (\ref{P1}), the optimal solution of the STBC-NJD scheme is hard to obtain. Instead, we consider a sub-optimal solution of the STBC-JD scheme, where the time allocation is obtained by solving (\ref{P1}) under the assumption that the DN only decodes users' messages in the $4$-th time slot, i.e., the STBC-NJD scheme. The solution can be efficiently obtained using Algorithm 1, and denoted by $\mathbf{\bar{t}} = \left[\bar{t}_1,\bar{t}_2,\bar{t}_3,\bar{t}_4\right]'$. Accordingly, the following throughput is achievable when the DN has joint decoding capability:
\begin{equation}
\bar{R} = \min \left( \bar{R}_X^{(2)},\bar{R}_Y^{(3)},\bar{R}_X^{(4)},\bar{R}_Y^{(4)}\right),
\end{equation}
where
\begin{align}
\bar{R}_X^{(2)} &= \bar{t}_2\log_{2}\left(1+ \rho_1 \frac{\bar{t}_1}{\bar{t}_2+\bar{t}_4}\right),\ \bar{R}_Y^{(3)} =\bar{t}_3\log_{2}\left(1+ \rho_2 \frac{\bar{t}_1}{\bar{t}_3+ \bar{t}_4}\right) ,   \\
     \bar{R}_X^{(4)}&= \bar{t}_4/2\log_{2}\left(1+ \rho_3 \frac{\bar{t}_1}{\bar{t}_2+\bar{t}_4}+ \rho_4 \frac{\bar{t}_1}{\bar{t}_3+\bar{t}_4}\right)
       +\bar{t}_2\log_{2}\left(1+ \rho_3 \frac{\bar{t}_1}{\bar{t}_2+\bar{t}_4}\right),\\
     \bar{R}_Y^{(4)}&= \bar{t}_4/2\log_{2}\left(1+ \rho_3 \frac{\bar{t}_1}{\bar{t}_2+\bar{t}_4}+ \rho_4 \frac{\bar{t}_1}{\bar{t}_3+\bar{t}_4}\right)
       +\bar{t}_3\log_{2}\left(1+ \rho_4 \frac{\bar{t}_1}{\bar{t}_3+\bar{t}_4}\right).
\end{align}

\section{Throughput Performance of Coherent User Cooperation}
In this section, we continue to study the throughput performance of the proposed user cooperation scheme when CSIT is available at each user, such that the two users can transmit coherently to the DN to further enhance the communication performance. In particular, we propose efficient algorithms to obtain the optimal throughput performance under both DTB-NJD and DTB-JD schemes. It is worth noting that the optimal throughput performance of the DTB-JD method also achieves the capacity of the proposed user cooperation method in Fig.~1. This is because the DTB-based cooperation maximizes the receive SNR at the DN and the joint-decoding scheme is capacity-achieving given the transmission method.

\subsection{Optimal Throughput of the DTB-NJD Scheme}
A major difference between the DTB-based and the STBC-based cooperation methods is the transmission time for the two users' messages in the $4$-th time slot, i.e., $t_4^{(1)}$ and $t_4^{(2)}$. Unlike the equal time allocation $t_4^{(1)}= t_4^{(2)}$ for the STBC-based cooperation methods, $t_4^{(1)}$ and $t_4^{(2)}$ can be different for the two DTB-based cooperation methods.

\begin{figure}
  \centering
   \begin{center}
      \includegraphics[width=0.65\textwidth]{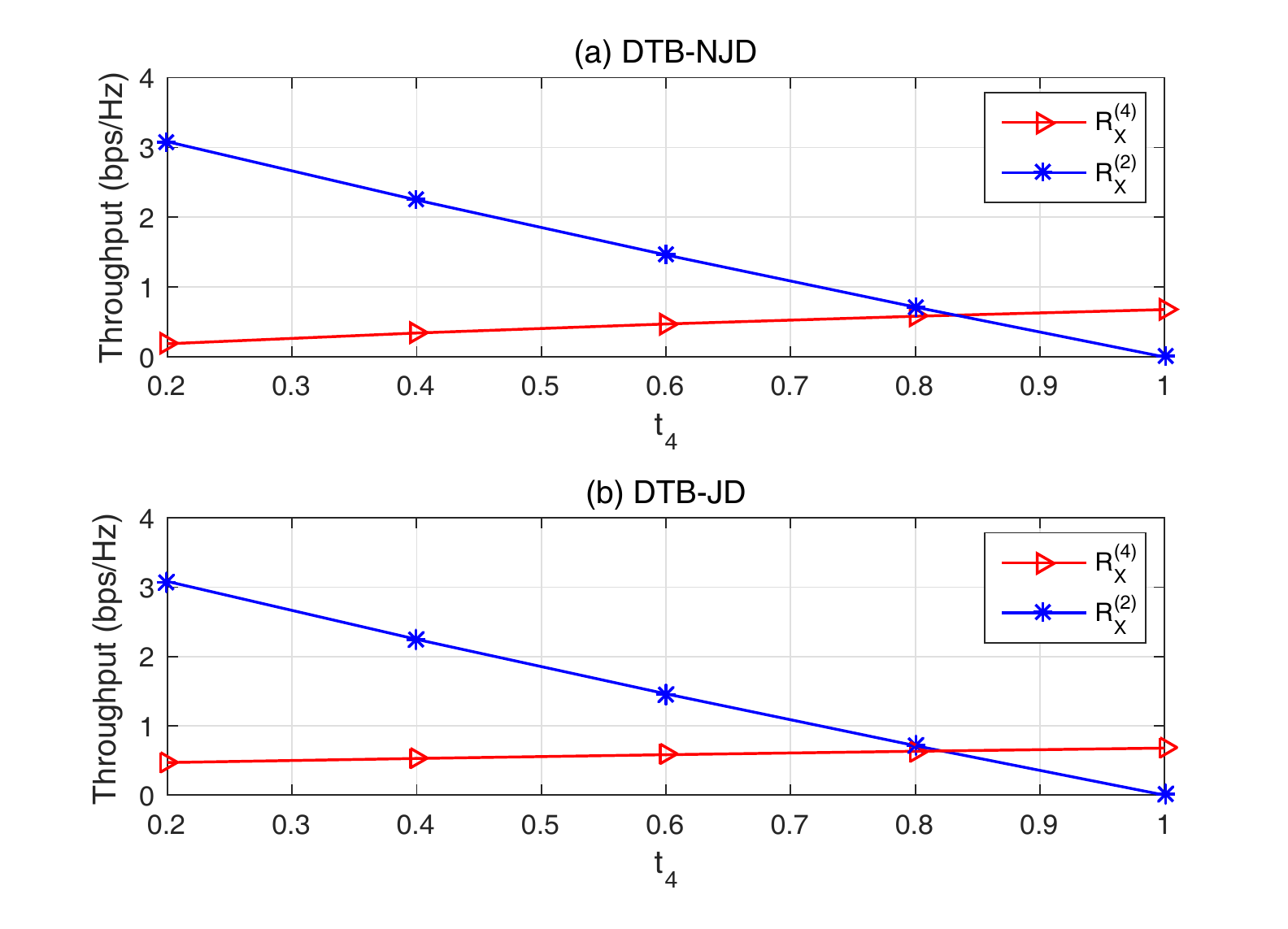}
   \end{center}
  \caption{Numerical results of the monotonic properties of DTB-based user cooperation.}
  \label{Fig.mrt}
\end{figure}

We first study the optimal throughput performance of the DTB-NJD scheme. Recall that the achievable data rates of X and Y are
\begin{subequations}
\label{15}
\begin{align}
R_X^{(2)}&=t_2\log_{2}\left(1+ \beta_1 \frac{t_1}{t_2+t_4}\right),\ \ R_Y^{(3)}=t_3\log_{2}\left(1+ \beta_2 \frac{t_1}{t_3+t_4}\right),\label{14}\\
R_X^{(4)}&=t_4^{(1)}\log_{2} \left(1+{\left({\sqrt{\beta_3 \frac{t_1}{t_2+t_4}} + \sqrt{\beta_4 \frac{t_1}{t_3+t_4}}}\right)^2} \right),\label{11}\\
R_Y^{(4)}&=t_4^{(2)}\log_{2} \left(1+{\left({\sqrt{\beta_3 \frac{t_1}{t_2+t_4}} + \sqrt{\beta_4 \frac{t_1}{t_3+t_4}}}\right)^2} \right), \label{12}
\end{align}
\end{subequations}
where ${\beta_1}\triangleq{ \eta P_t h_{EX} h_{XY} } / {N_0}$, ${\beta_2}\triangleq{ \eta P_t h_{EY} h_{YX} } / {N_0}$, ${\beta_3}\triangleq{ \eta P_t h_{EX} h_{XD}}/ {N_0}$, and ${\beta_4}\triangleq{ \eta P_t h_{EY} h_{YD}}/ {N_0}$.
Following the similar analysis in Section IV.A, we can easily see that the optimal solution must satisfy
\begin{equation}\label{P3}
  R_X^{(2)*}=R_Y^{(3)*}=R_X^{(4)*}=R_Y^{(4)*}.
\end{equation}
Evidently, we can infer from (\ref{11}) and (\ref{12}) that $t_4^{(1)} = t_4^{(2)}=t_4/2$ always holds for the optimal solution. Besides, following Lemma 4.1, we can see that, given fixed $t_1$ and $t_4$, $R_X^{(2)}$ ($R_Y^{(3)}$) increases (decreases) with $t_2 \in \left(1- t_1-t_4\right)$. In addition, it also holds that, given a fixed $t_1$, $R_X^{(2)}$ (and $R_Y^{(3)}$ ) decreases and $R_X^{(4)}$ ($R_Y^{(4)}$) increases in $t_4 \in\left[1-t_1\right]$ to satisfy $R_X^{(2)} = R_Y^{(3)}$, whose proof is omitted to avoid repetition and demonstrated numerically in Fig.~\ref{Fig.mrt}(a). The monotonic properties of $\left\{R_X^{(2)},R_Y^{(3)},R_X^{(4)},R_Y^{(4)}\right\}$ with respect to $\left(t_2,t_3,t_4\right)$ as well as the optimality conditions are exactly the same for the DTB-NJD and STBC-NJD schemes, thus the optimal time allocation of the DTB-NJD scheme can also be solved with Algorithm 1, only with the replacement of the expressions of $\left\{R_X^{(2)},R_Y^{(3)},R_X^{(4)},R_Y^{(4)}\right\}$ in Algorithm 1 by those in (\ref{15}), and $t_4^{(1)} = t_4^{(2)}=t_4/2$ in (\ref{11}) and (\ref{12}).

\subsection{Optimal Throughput of the DTB-JD Scheme}
For the DTB-JD scheme, we can infer that $R_X^{(2)*}=R_Y^{(3)*}=R_X^{(4)*}=R_Y^{(4)*}$ also holds for the optimal solution, where the rate expressions are
\begin{subequations}
\label{16}
\begin{align}
R_X^{(2)}&=t_2\log_{2}\left(1+ \beta_2 \frac{t_1}{t_2+t_4}\right),\ \ R_Y^{(3)}=t_3\log_{2}\left(1+ \beta_2 \frac{t_1}{t_3+t_4}\right),\label{17}\\
R_X^{(4)}& = t_4^{(1)}\log_{2} \left(1+{\left({\sqrt{\beta_3 \frac{t_1}{t_2+t_4}} + \sqrt{\beta_4 \frac{t_1}{t_3+t_4}}}\right)^2} \right)
       + t_2\log_{2}\left(1+\beta_3 \frac{t_1}{t_2+t_4} \right ), \label{18}\\
R_Y^{(4)} & = t_4^{(2)}\log_{2} \left(1+{\left({\sqrt{\beta_3 \frac{t_1}{t_2+t_4}} + \sqrt{\beta_4 \frac{t_1}{t_3+t_4}}}\right)^2} \right)
       + t_3\log_{2}\left(1+\beta_4 \frac{t_1}{t_3+t_4}\right). \label{19}
\end{align}
\end{subequations}
In this case, however, $t_4^{(1)}$ and $t_4^{(2)}$ are not equal for the optimal solution in general, which is the main difference between the DTB-JD and other three schemes. As we discussed before, We can always find a unique set of $(t_2,t_3)$ to satisfy $R_X^{(2)} = R_{Y}^{(3)}$ when a set of $(t_1,t_4)$ is given by a bi-section search over $t_2\in\left[0,1-t_0-t_1-t_4\right]$. After that, a unique set of $(t_4^{(1)},t_4^{(2)})$ can be found to satisfy $R_X^{(4)} = R_{Y}^{(4)}$ by a bi-section search over $t_4^{(1)}\in\left[0,t_4 \right]$ when a set of $(t_1,t_2,t_3,t_4)$ is given. This is because, $R_X^{(4)}$ and $R_Y^{(4)}$ can be denoted as functions of $t_4^{(1)}$ and $t_4^{(2)}$, respectively, where $t_4^{(1)}+t_4^{(2)}=t_4$ is a fixed parameter. Accordingly, we can calculate $R_X^{(2)}$ and $R_X^{(4)}$ using (\ref{17}) and (\ref{18}), based on which we can find a unique $t_4$ that satisfy $R_X^{(2)}(t_1,t_4) = R_X^{(4)}(t_1,t_4)$ using a bi-section search over $t_4\in[0,1-t_0-t_1]$. Then, we only need to perform a linear search over $t_1\in[0,1-t_0]$ to find the optimal set of $(t_2,t_3,t_4^{(1)},t_4^{(2)})$ that achieves the largest common throughput. In particular, the monotonic properties of $R_X^{(2)}(t_1,t_4)$ and $R_X^{(4)}(t_1,t_4)$ as a function of $t_4$ can be proved similarly as Lemma 4.2 in Appendix B and illustrated numerically in Fig.~\ref{Fig.mrt}(b).  A pseudo-code of the above searching algorithm is summarized in Algorithm 2, whose time complexity is $1/\Delta \cdot \left[\log(1/\sigma)\right]^3$.

\begin{algorithm}
\footnotesize
 \SetAlgoLined
 \SetKwData{Left}{left}\SetKwData{This}{this}\SetKwData{Up}{up}
 \SetKwRepeat{doWhile}{do}{while}
 \SetKwFunction{Union}{Union}\SetKwFunction{FindCompress}{FindCompress}
 \SetKwInOut{Input}{input}\SetKwInOut{Output}{output}
 \Input{time duration $T=1$, channel estimation time $t_0$}
 \Output{the optimal time allocation of $\left\{t_1^*,t_2^*,t_3^*,t_4^{(1)*},t_4^{(2)*}\right\}$}
 Initialize: $t_1\leftarrow 0$, $R^*\leftarrow 0$, $\Delta \leftarrow$ small positive step size\;
   \While{$t_1 \leq 1-t_0$}{
 $t_1 \leftarrow t_1+\Delta$\;
    $UB_4 \leftarrow  1-t_0-t_1$, $LB_4 \leftarrow 0$\;
    \Repeat{$|R_X^{(2)}-R_X^{(4)}|<\sigma$}{
    $t_4\leftarrow \left(UB_4+LB_4\right)/2$\;
    $UB_2 \leftarrow  1 - t_0 - t_1 -t_4$, $LB_2 \leftarrow 0$\;
    \Repeat{$|R_X^{(2)}-R_Y^{(3)}|<\sigma$}{
    $t_2\leftarrow \left(UB_2+LB_2\right)/2$\;
    $t_3\leftarrow 1-t_0-t_1-t_4-t_2$\;
    Calculate $R_X^{(2)}$ and $R_Y^{(3)}$ using (\ref{17})\;
    \eIf{$R_X^{(2)}>R_Y^{(3)}$}{
    $UB_2 \leftarrow t_2$;}
    {
    $LB_2 \leftarrow t_2$;
    }
    }

    $UB_5 \leftarrow  t_4$, $LB_5 \leftarrow 0$\;
    \Repeat{$|R_X^{(4)}-R_Y^{(4)}|<\sigma$}{
    $t_4^{(1)}\leftarrow \left(UB_5+LB_5\right)/2$\;
    $t_4^{(2)}\leftarrow t_4-t_4^{(1)}$\;
    Calculate $R_X^{(4)}$ and $R_Y^{(4)}$ using (\ref{18}) and (\ref{19}), respectively\;
    \eIf{$R_X^{(4)}>R_Y^{(4)}$}{
    $UB_5 \leftarrow t_4^{(1)}$;}
    {
    $LB_5 \leftarrow t_4^{(1)}$;
    }
    }

    \eIf{$R_X^{(4)}>R_X^{(2)}$}{
    $UB_4 \leftarrow t_4$;}
    {
    $LB_4 \leftarrow t_4$;
    }
    }
    $R\leftarrow \min\left(R_X^{(2)},R_X^{(4)}\right)$\;
    \If{$R>R^*$}{
    $R^*\leftarrow R$, \ \
    $\left\{t_1^*,t_2^*,t_3^*,t_4^{(1)*},t_4^{(2)*}\right\}\leftarrow \left\{t_1,t_2,t_3,t_4^{(1)}, t_4^{(2)} \right\}$\;
    }

}
\textbf{Return} $\left\{t_1^*,t_2^*,t_3^*,t_4^{(1)*},t_4^{(2)*}\right\}$.
\caption{Optimal solution to (\ref{P1}) for the DTB-JD scheme.}
\label{alg2}
\end{algorithm}

\section{Benchmark Methods}
\begin{figure}
  \centering
   \begin{center}
      \includegraphics[width=0.9\textwidth]{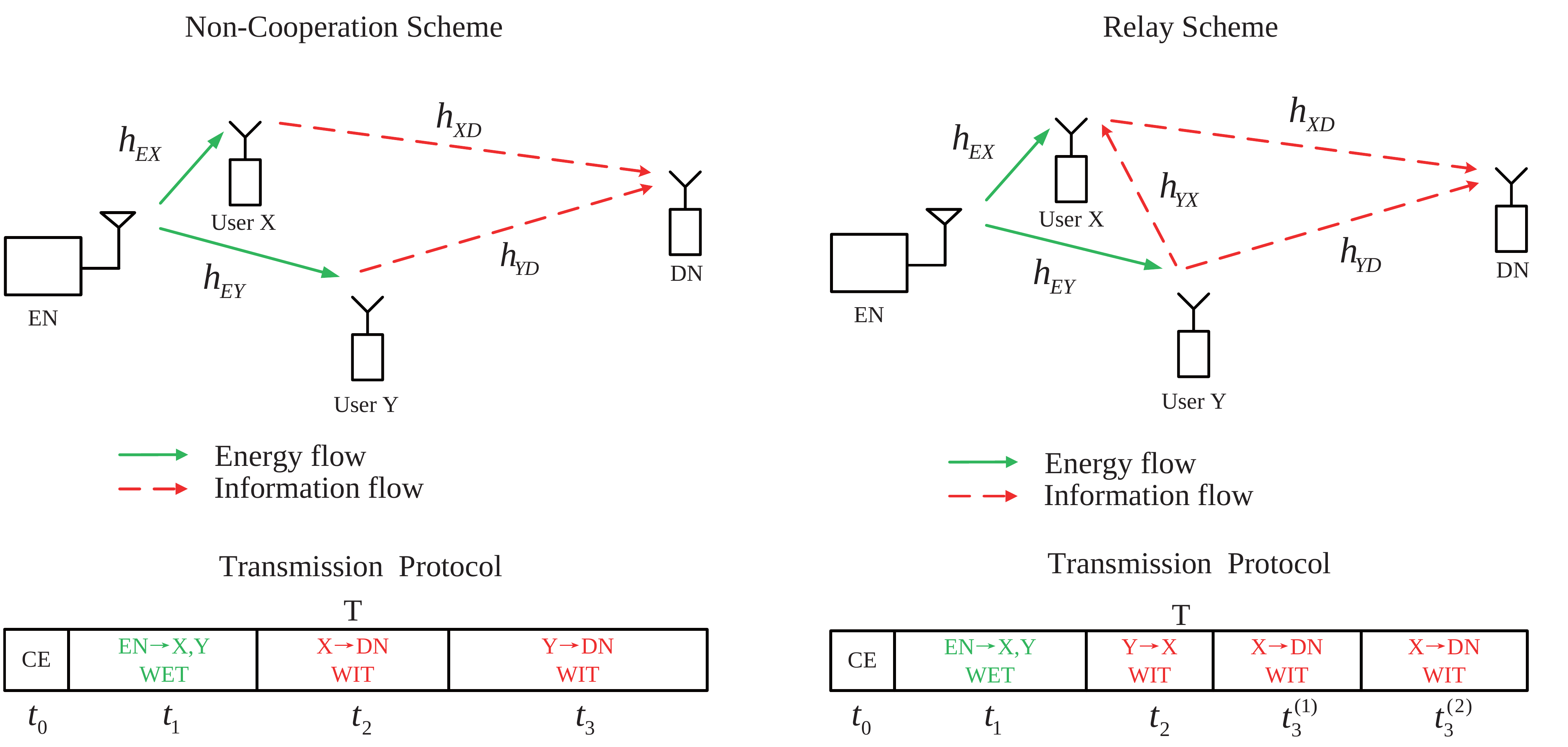}
   \end{center}
  \caption{The considered benchmark methods}
  \label{Fig.5}
\end{figure}
For performance comparison, we consider in Fig.~\ref{Fig.5} two benchmark methods: the two users do not cooperate and transmit to the DN in a TDMA manner (Non-cooperation); and one user acts as the relay for the other (Relay) \cite{2014:Ju3,2015:HeCHen,2015:Nasir,2008:Z.Zhou,2003:Laneman,2004:Laneman,2007:Bletsas}. For both methods, CE consumes the same amount of time $t_0$, the first time slot $t_1$ is assigned for WET and the remaining time is used for WIT.
\subsection{Cooperation by Relaying}
For the Relay scheme, the WIT time is divided into two time slots $t_2$ and $t_3$. During $t_2$, one user uses the harvested energy to transmit its own information to the other. In $t_3$, the other user will help forward the information received in $t_2$ and transmit its own information to the DN. In particular, either user can act as the relay for the other (i.e., $Y{\rightarrow}X{\rightarrow}D$ or $X{\rightarrow}Y{\rightarrow}D$) depending on the channel conditions. In this paper, we choose the better one between the two scenarios that yields higher throughput to represent the relay scheme under different channel conditions. When user $X$ acts as the relay for user $Y$(i.e., $Y{\rightarrow}X{\rightarrow}D$) and the DN does not have joint decoding capability (i.e., Relay-NJD), we can infer that the achievable rates are
\begin{subequations}
\label{20}
\begin{align}
R_Y^{(2)}&=t_2\log_{2}\left(1+\frac{E_Yh_{YX}}{t_2N_0}\right),\label{21}\\ R_Y^{(3)}&=t_3^{(1)}\log_{2}\left(1+\frac{E_Xh_{XD}}{t_3N_0}\right),\label{22}\\
R_X^{(3)}&=t_3^{(2)}\log_{2}\left(1+\frac{E_Xh_{XD}}{t_3N_0}\right). \label{23}
\end{align}
\end{subequations}
Otherwise, when the decoder of the DN can decode the information received during $t_2$ (i.e., Relay-JD), then (\ref{22}) can be repalaced by
\begin{equation}
\label{24}
R_Y^{(3)}=t_3^{(1)}\log_{2}\left(1+\frac{E_Xh_{XD}}{t_3N_0}\right)+t_2\log_{2}\left(1+\frac{E_Yh_{YD}}{t_2N_0}\right).
\end{equation}
From the above discussions, the achievable rates of user $X$ and $Y$ are
\begin{equation}
  R_X = R_X^{(3)},\ R_Y = \min \left(R_Y^{(2)},R_Y^{(3)}\right),
\end{equation}
where $R_Y^{(2)}$ and $R_X^{(3)}$ are in (\ref{21}) and (\ref{23}), respectively. For $R_Y^{(3)}$, its expression depends on the specific receiver structures with which $R_Y^{(3)}$ is (\ref{22}) or (\ref{24}).

\subsection{Non-cooperating Users}
Different from user cooperation and relay scheme, user $X$ and $Y$ of the non-cooperation scheme transmit their independent information to the DN directly in $t_2$ and $t_3$, respectively, and the achievable rates of user $X$ and $Y$ are
\begin{equation}
\label{rx2non}
R_X^{(2)}=t_2\log_{2}\left(1+\frac{E_Xh_{XD}}{t_2N_0}\right)£¬
\end{equation}
\begin{equation}
\label{ry3non}
R_Y^{(3)}=t_3\log_{2}\left(1+\frac{E_Yh_{YD}}{t_3N_0}\right).
\end{equation}
In this case, the achievable rates of user $X$ and $Y$ can be denoted as
\begin{equation}
  R_X = R_X^{(2)},\ R_Y =R_Y^{(3)} ,
\end{equation}
where $R_X^{(2)}$ and $R_Y^{(3)}$ are in (\ref{rx2non}) and (\ref{ry3non}), respectively.

Both two benchmark methods adopt the following optimal time allocation problem to maximize the throughput
\begin{equation}
   \begin{aligned}
    &\max_{t_1,t_2,t_3} & &  \min(R_X,R_Y)\\
    &\text{s. t.}    & & t_1+t_2+t_3 = 1 - t_0, \label{P1non}\\
    & & & t_1,t_2,t_3\geq 0,
   \end{aligned}
\end{equation}
and base on which we can obtain the optimal time slot allocation using some search methods (i.e., bi-section search or line search), which are omitted due to the scope of this paper.

\section{Simulation Results}
In this section, we evaluate the performance of the proposed user cooperation under different channel conditions. In all figures, the optimal throughput performance of different schemes are presented. Unless otherwise stated, it is assumed that the distance between the EN and user $X$ and $Y$ is $5$m and $10$m, respectively, and the users are separated by $2$m. We consider using Powercast TX91501-3W power transmitter at the EN and P2110 Powerharvester receiver at the users. \footnote{Please refer to the website of Powercast Corp. (http://www.powercastco.com) for detailed product specifications.} In this case, the transmit power of EN is $P_t=3$W, and the wireless channel gain $h_{ij}=G_A(\frac{3\cdot10^8}{4{\pi}df_d})^{d_{D}}$ \cite{2005:Goldsmith}, where $ij\in\{EX,EY,XY,YX,XD,YD\}$, $f_d$ denotes $915$MHz carrier frequency, $d_{D} = 2$ denotes the path loss exponent, and the antenna power gain $G_A=2$.

\begin{figure}
  \centering
   \begin{center}
      \includegraphics[width=0.65\textwidth]{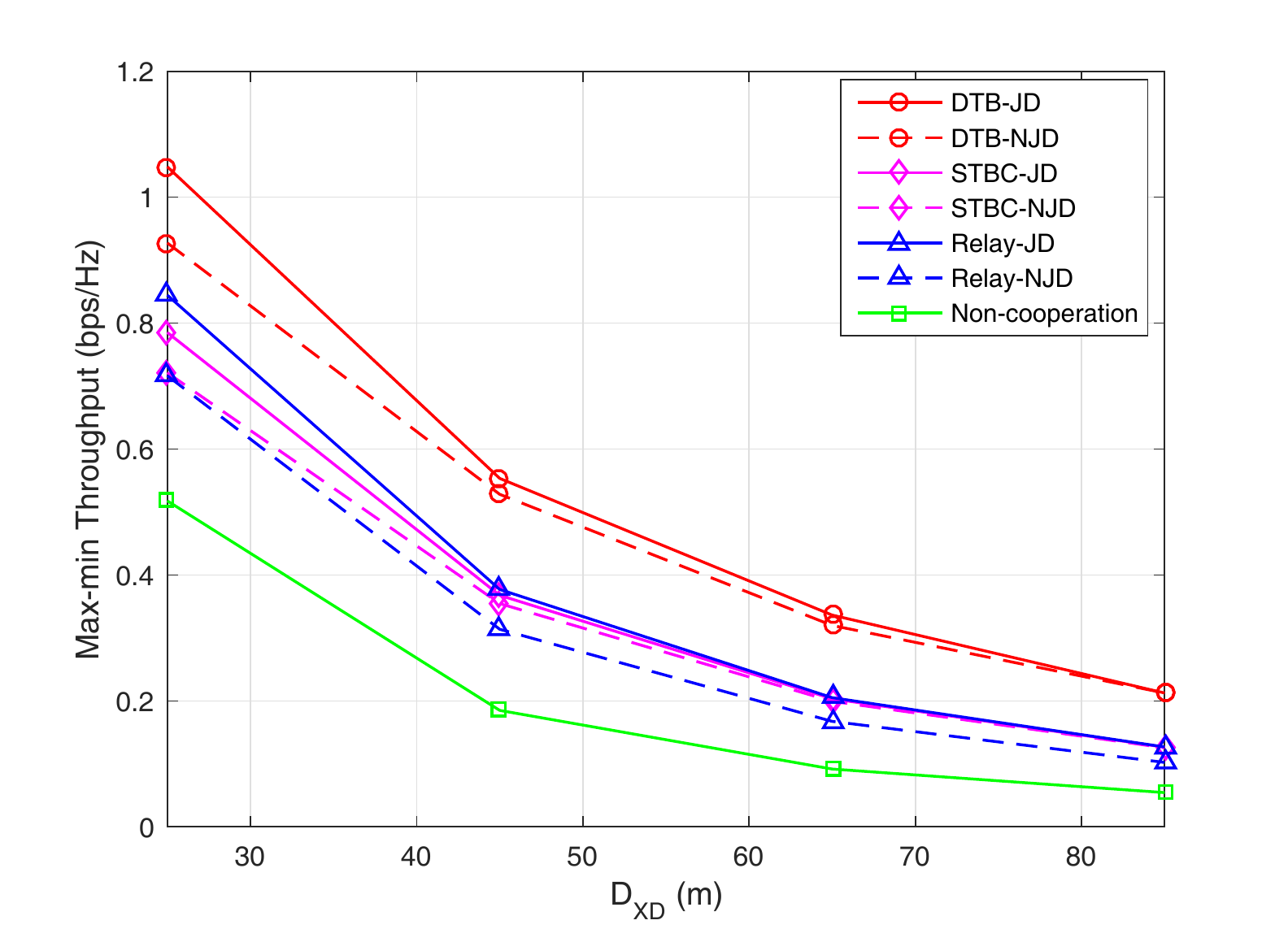}
   \end{center}
  \caption{The impact of user-to-DN channel to the optimal common throughput performance.}
  \label{Fig.6}
\end{figure}

Fig.~\ref{Fig.6} shows the impact of user-to-DN channel to the optimal common throughput performance. Here, we set $h_{XD}=h_{YD}$ and $h_{EX}=4h_{EY}=2.72\times10^{-5}$, and vary the distance between two users and DN from $25$m to $85$m. It is observed that all schemes decrease as the user-to-DN channels degrade. In particular, the proposed user cooperation method with DTB-based transmission has the best performance among all the schemes considered. The STBC-based cooperation scheme perform closely with the relay scheme. The non-cooperation scheme has the worst performance because of its inability to address the problem of unbalanced energy harvested by the two users. Note that joint decoding at the DN can significantly improve the throughput performance of NJD schemes when the user-to-DN channels are strong. However, the improvement becomes marginal as the user-to-DN channels become very weak, e.g., separated by over $50$ meters. This is because the cooperation time, and thus the time duration that the DN overhears, is much shorter than the direct information transmission from user to the DN. Fig.~\ref{Fig.6} shows that the proposed cooperation method is robust against user-to-DN channel degradation.

\begin{figure}
  \centering
   \begin{center}
      \includegraphics[width=0.65\textwidth]{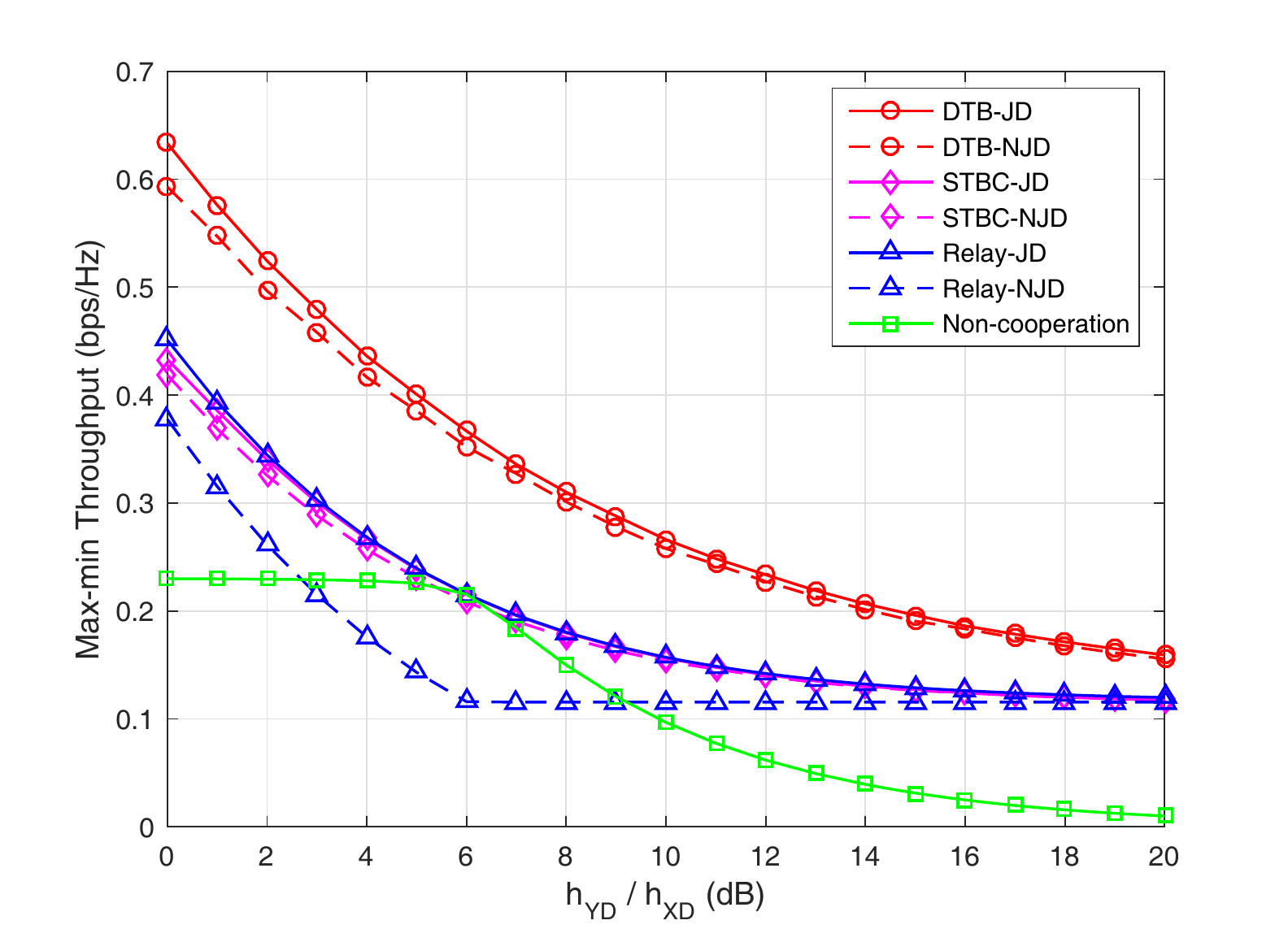}
   \end{center}
  \caption{The impact of user-to-DN channel disparity to the optimal common throughput performance.}
  \label{Fig.7}
\end{figure}

Fig.~\ref{Fig.7} shows the impact of user-to-DN channel disparity to the optimal common throughput performance. Without loss of generality, we fix $h_{YD}=4.25\times10^{-7}$ (this corresponds to the average channel gain when $Y$ is $40$ meters from the DN) as a constant and show the performance when $h_{XD}$ becomes smaller. Note that when $h_{YD}/h_{XD}$ changes from $0$ to $5 $dB, the common throughput of the non-cooperation scheme hardly changes while those of the cooperative scheme and relay scheme decrease more evidently. This is because the $0-5$ dB case corresponds to the energy-constrained region, where the major performance bottleneck is the less energy harvested by $Y$ due to the poor EN-to-Y channel. Therefore, moderate decrease of user $X$'s data rate will not change the common throughput performance of the non-cooperation scheme. For the Relay and user cooperation schemes, however, the data rate performance is more sensitive to the channel degradation of X-to-DN channel, as it needs to transmit the messages of both two users.
Obviously, the proposed DTB-based methods outperform the other schemes regardless of the joint decoding capability. The proposed STBC-based cooperation methods and the relay scheme perform similarly when joint decoding capability is achievable. However, when joint decoding capability is not achievable at the DN, the relay scheme performs poorly, where we can observe an evident switch from user $X$ being the relay to user $Y$ being the relay when $h_{YD}/h_{XD} > 6$ dB. As we further decrease the channel gain of $h_{XD}$, we can see that the performance of the proposed cooperation scheme gradually approaches that of the Relay scheme, as now most data is sent from user Y to the DN. Fig.~\ref{Fig.7} shows that the proposed cooperation method is robust against user-to-DN channel disparity under different transmission schemes because of the channel diversity achieved in transmitting user messages.

\begin{figure}
  \centering
   \begin{center}
      \includegraphics[width=0.65\textwidth]{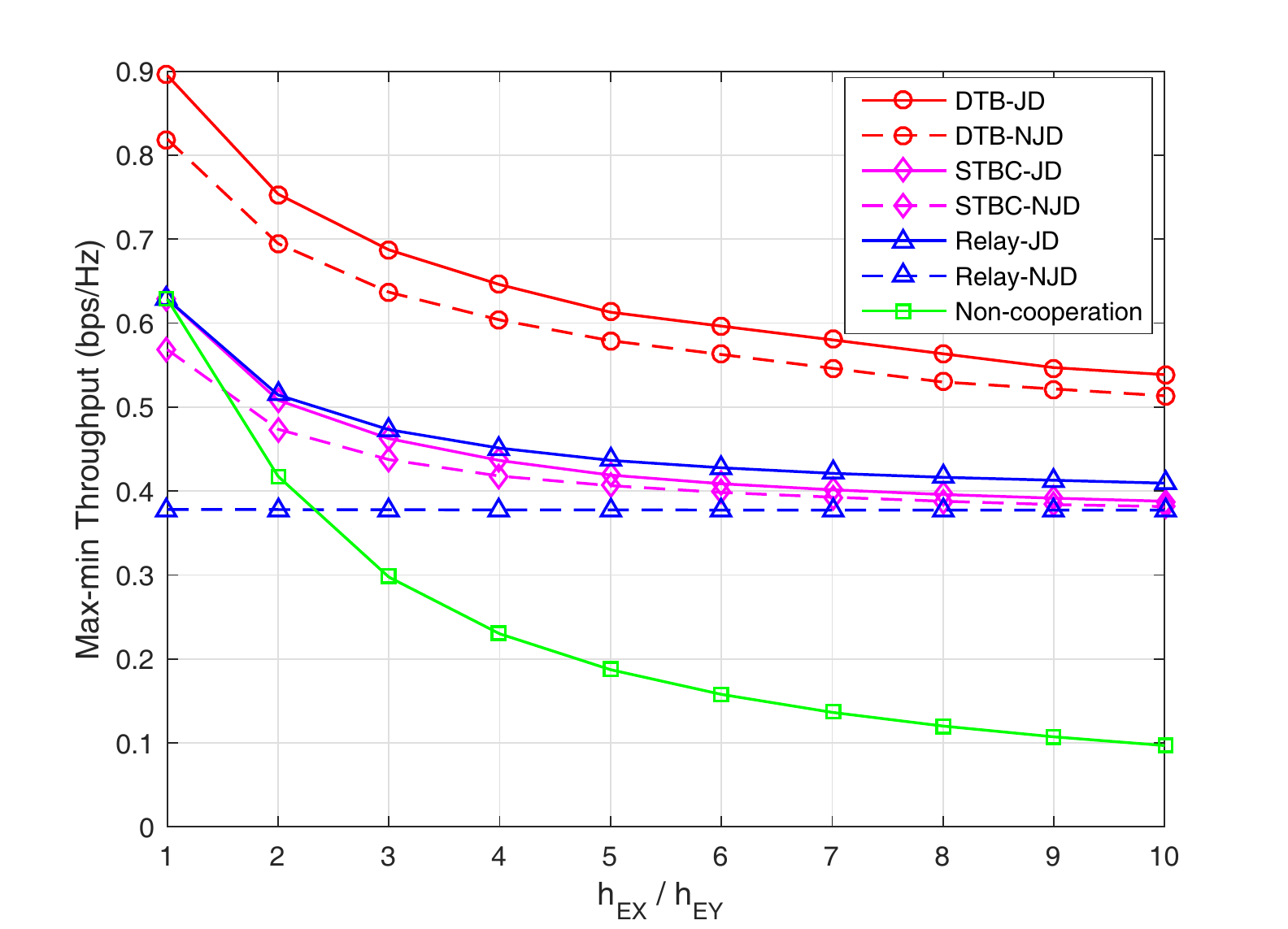}
   \end{center}
  \caption{The impact of EN-to-user channel disparity to the optimal common throughput performance.}
  \label{Fig.8}
\end{figure}

Fig.~\ref{Fig.8} shows the impact of EN-to-user channel disparity to the optimal common throughput performance. Here, we set $h_{XD}=h_{YD}=4.25\times10^{-7}$, fix $h_{EX}=2.72\times10^{-5}$ as a constant and show the performance when $h_{EY}$ becomes smaller. Notice that the performance of non-cooperation scheme degrades significantly as $h_{EY}$ becomes smaller, while the proposed cooperation scheme degrades moderately. It is worth noting that the performance of the Relay-NJD scheme ($Y{\rightarrow}X{\rightarrow}D$) changes marginally compared other schemes as $h_{EY}$ changes. This is because the distance between two users is very short so that moderate decrease of user $Y$'s harvested energy has marginal impact on the throughput. The performance of the DTB-based cooperation scheme has evident advantages over the relay scheme and non-cooperation scheme. The results in Fig.~\ref{Fig.8} demonstrate the superior performance of the proposed user cooperation method, thanks to the energy diversity achieved from allowing the users to share their energy to transmit jointly their messages.

\begin{figure}
  \centering
   \begin{center}
      \includegraphics[width=0.65\textwidth]{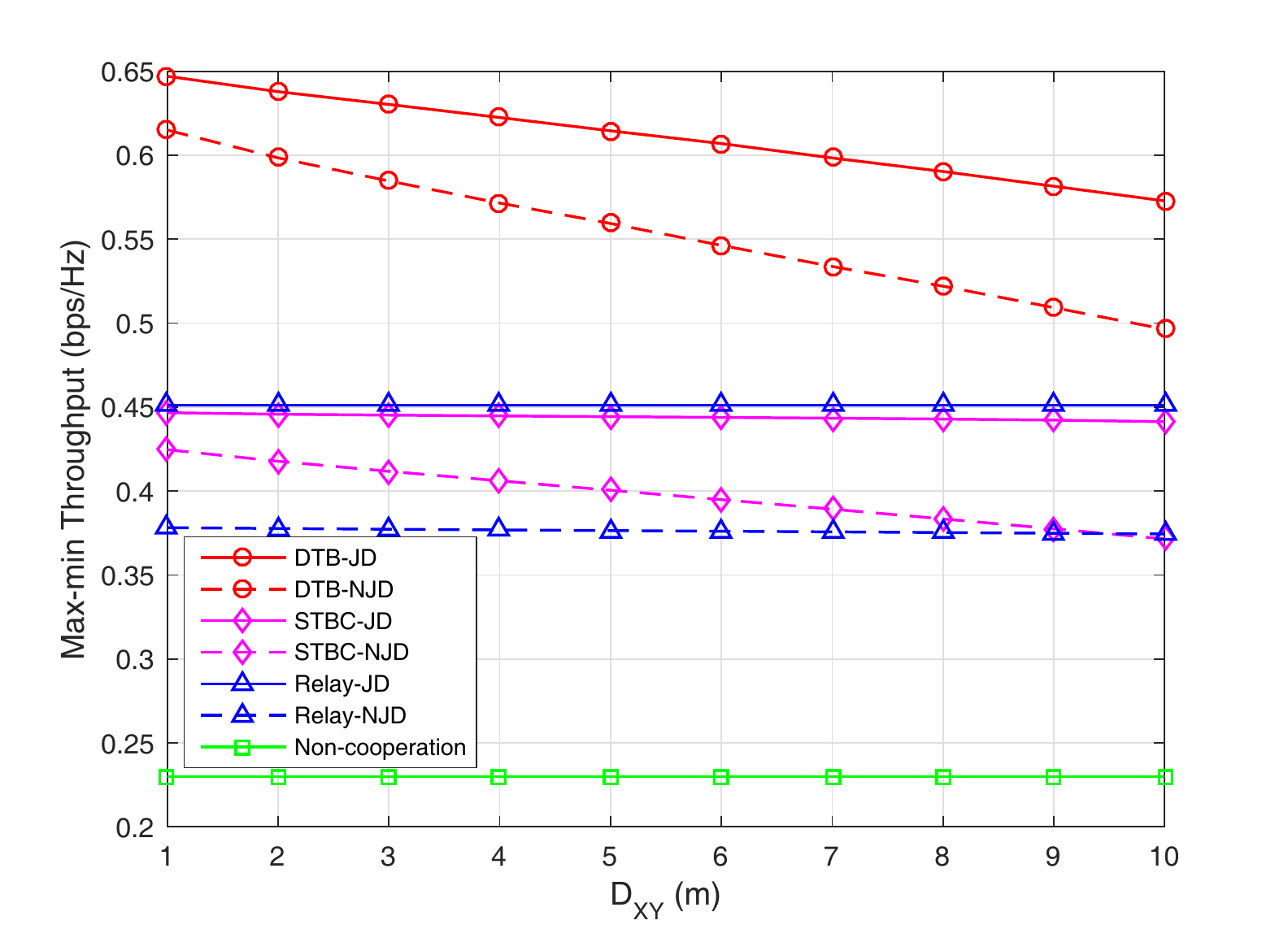}
   \end{center}
  \caption{The impact of inter-user channel strength to the optimal common throughput performance.}
  \label{Fig.9}
\end{figure}

In addition, Fig.~\ref{Fig.9} shows the impact of inter-user channel strength to the optimal common throughput performance. Here, we set $h_{XD}=h_{YD}=4.25\times10^{-7}$ and $h_{EX}=4h_{EY}=2.72\times10^{-5}$, and vary the distance between user $X$ and $Y$ from $1$m to $10$m. As the performance of non-cooperation scheme is independent of $D_{XY}$, its throughput does not change with the inter-user channel conditions. It is observed that the max-min throughput of cooperation and relay scheme decreases with $D_{XY}$. However, the cooperation scheme is more sensitive to the channel degradation between the cooperating users than the relay scheme because it uses the inter-user channel twice during the information exchange while the relay scheme only needs once. We can therefore conclude that user cooperation is most effective when the inter-user channel is sufficiently strong to support efficient user message exchange.

It is also worth noting we do not intend to claim that the proposed user cooperation method has the best performance in all scenarios. In general, different scheme should be applied based on the network setups and parameters. However, the proposed user cooperation method has demonstrated superior performance under different setups, especially when two users are close with each other so that the inter-user channel is good enough and the two users have similar user-to-DN channels. In practice, this includes extensive application scenarios, such as IoT/IoE systems or WSNs for environment monitoring, etc, where neighboring low-power wireless devices are often close to each other (strong inter-user channel), and far-away to the information collection point (the user-to-DN channels are mostly comparable). In particular, coherent cooperation methods (i.e, DTB-based cooperation methods), have evident performance gain over the other methods, which shows the importance of the availability of CSI knowledge to the system performance. In the case of non-joint decoding DN, non-coherent cooperation shows robust and superior performance than relay method under different scenarios. It is also effective to improve the performance by using joint decoding capability, especially when user-to-DN channel condition is good. Thanks to the channel and energy diversity gains achieved, the proposed cooperation method shows robust performance under most scenarios in either coherent or non-coherent manner.

\section{Conclusion}
This paper studied a two-user WPCN in which a new user cooperation method is exploited to improve the throughput fairness. For users using both coherent and non-coherent cooperations, we derived the maximum common throughput achieved by the proposed user cooperation and performed numerical analysis to study the impact of system setups to the throughput performance. By comparing with two representative benchmark methods, we showed that the proposed user cooperation method can effectively achieve both channel and energy diversity gains to enhance the throughput fairness under different setups, especially when the inter-user channels are sufficiently strong to support efficient information exchange between the two users, and the two users have similar user-to-DN channels. In particular, the proposed DTB-JD scheme achieves the capacity under the considered user cooperation protocol, which has evident performance gain over the other methods.

\appendices
\section{Proof of Lemma 4.1}
The transmit power of user $X$ is $P_X=E_X/(t_2+t_4)$, we have from (\ref{Rx2}) that
\begin{align}\label{proof3.1}
R_X^{(2)}&= t_2\log_{2}\left(1+\frac{E_Xh_{XY}}{(t_2+t_4)N_0}\right) \triangleq t_2\log_{2}\left(1+ \frac{c_1}{t_2+c_2}\right),
\end{align}
where $c_1\triangleq E_Xh_{XY}/N_0$, $c_2 \triangleq t_4$ are both constant. By taking the first and second order derivatives of $R_X^{(2)}$ in $t_2$, we have
\begin{align}\label{}
\frac{dR_X^{(2)}}{dt_2} & =\log_{2}\left(1+ \frac{c_1}{t_2+c_2}\right)- \frac{c_1t_2}{\ln2(t_2+c_3)(t_2+c_2)}, \\
\frac{d^2R_X^{(2)}}{dt_2^2} & = - \frac{c_1}{\ln2}  \frac{(c_2+c_3)t_2+2c_2c_3}  {(t_2+c_3)(t_2+c_2)},
\end{align}
where $c_3\triangleq c_1+c_2$. Because $\frac{d^2R_X^{(2)}}{dt_2^2}<0$ and $ \lim\limits_{t_2 \to +\infty} \frac{dR_X^{(2)}}{dt_2} = 0 $, we can infer that $ \frac{dR_X^{(2)}}{dt_2} > 0 $ when $t_2>0$, which leads to the proof of Lemma 4.1 that $R_X^{(2)}$ increases in $t_2 \in\left[0,T_0\right]$. Similarly, we have $R_Y^{(3)}$ deceases with $t_2 \in \left[0,T_0\right]$.  $\hfill \blacksquare$

\section{Proof of Lemma 4.2}
First of all, we show that both $t_2$ and $t_3$ decrease as $t_4$ increases. Otherwise, we assume without loss of generality that $t_2$ increases and $t_3$ decreases when $t_4$ become larger. We denote the updated values of $t_2$ and $t_3$ after $t_4$ becomes $\bar{t}_4=t_4+{\Delta}t_4$ as $\bar{t}_2=t_2+{\Delta} t_2$ and $\bar{t}_3=t_3-{\Delta} t_3$, respectively, where ${\Delta} t_2,{\Delta} t_3,{\Delta}t_4>0$, and ${\Delta} t_2+ {\Delta}t_4 -{\Delta} t_3 =0$. Besides, we denote the updated values of $R_X^{(2)}$ and $R_Y^{(3)}$ as $\bar{R}_X^{{(2)}}$ and $\bar{R}_Y^{{(3)}}$, respectively. It can be easily shown from Lemma 4.1 that $\bar{R}_X^{{(2)}}>\bar{R}_Y^{{(3)}}$ given $R_X^{(2)} = R_Y^{(3)}$. However, this contradicts with the necessary condition of an optimal solution that requires $\bar{R}_X^{{(2)}}=\bar{R}_Y^{{(3)}}$. Therefore, we reject our assumption and conclude that both $t_2$ and $t_3$ decrease as $t_4$ increases. Because $t_2+ t_3 +t_4 =T_1$, we can infer that $t_2+t_4 = T_1 -t_3$ increases with $t_4$, so does $t_3+t_4$. This, together with the result that $t_2$ (and $t_3$) decrease with $t_4$, leads to the proof that $R_X^{(2)}$ in (\ref{rx2}) (and $R_Y^{(3)}$ in (\ref{ry3})) is a decreasing function with $t_4$.

Next, we show that $R_X^{(4)}$ in (\ref{rx4}) increases with $t_4$. To see this, we let $\bar{R}_X^{(4)}$ denote the updated value of $R_X^{(4)}$ after $t_4$ increases to $\bar{t}_4 = t_4 + \Delta t_4$. First, we can infer from $\Delta t_4 = \Delta t_2 + \Delta t_3$ and $\Delta t_2,\Delta t_3>0$ that $0< \Delta t_3 \leq \Delta t_4$ and $0<\Delta t_2\leq \Delta t_4$ hold. Then, we have
\begin{equation}
\begin{aligned}
\bar{R}_X^{(4)} &=  \frac{t_4+\Delta t_4}{2}\log_{2}\left( 1+ \rho_3 \frac{t_1}{1-t_1 - t_3 + \Delta t_3 }+ \rho_4 \frac{t_1}{1-t_1 - t_3 + \Delta t_2}\right)\\
&\geq \frac{t_4+\Delta t_4}{2}\log_{2}\left( 1+ \rho_3 \frac{t_1}{1-t_1 - t_3 + \Delta t_4 }+ \rho_4 \frac{t_1}{1-t_1 - t_3 + \Delta t_4}\right)\\
&\geq \frac{t_4}{2}\log_{2}\left( 1+ \rho_3 \frac{t_1}{1-t_1 - t_3 }+ \rho_4 \frac{t_1}{1-t_1 - t_3 }\right) = R_X^{(4)},
\end{aligned}
\end{equation}
where the first inequality holds because $0< \Delta t_3 \leq \Delta t_4$ and $0<\Delta t_2\leq \Delta t_4$, and the second inequality holds because $R_X^{(4)}$ increases monotonically with $t_4$ when $t_2$ and $t_3$ are fixed. This leads to the proof that $R_X^{(4)}$ increases with $t_4$. $\hfill \blacksquare$


%
%
%
%

\end{document}